%% file: drafts.tex
\pdfoutput=1

\documentclass{article}

\usepackage{spconf,amsmath,graphicx}
\usepackage[sort&compress, numbers]{natbib}
\usepackage[outdir=./]{epstopdf}

\def\imgsize{.30}
\def\imgs2{5.2 cm}

\def\correct<#1>{\textbf{* #1 *}}

\title{Objective Assessment of Spatial Audio Quality using Directional Loudness Maps}
\name{Pablo M. Delgado $^{\star}$, Jürgen Herre $^{\dagger \star}$ }

\address{$^{\star}$ International Audio Laboratories Erlangen $^{\ddagger}$, Am Wolfsmantel 33, 91058 Erlangen, Germany \protect\thanks{$^{\ddagger}$ A joint institution of the Friedrich-Alexander Universität Erlangen-Nürnberg (FAU) and Fraunhofer IIS, Germany.} \\
$^{\dagger}$ Fraunhofer IIS, Am Wolfsmantel 33, 91058 Erlangen, Germany \\
Correspondence should be addressed to pablo.delgado@audiolabs-erlangen.de\\
}
\begin{document}
%
\maketitle
\begin{abstract}


This work introduces a feature extracted from stereophonic/binaural audio signals
aiming to represent a measure of perceived quality degradation in processed spatial auditory scenes.
The feature extraction technique is based on a simplified stereo signal model considering auditory events positioned towards a given direction in the stereo field using amplitude panning (AP) techniques. We decompose the stereo signal into a set of directional signals for given AP values in the Short-Time Fourier Transform domain and calculate their overall loudness to form a directional loudness representation or maps.
Then, we compare directional loudness maps of a reference signal and a deteriorated version to derive a distortion measure aiming to describe the associated perceived degradation scores reported in listening tests. 
 
The measure is then tested on an extensive listening test database with stereo signals processed by state-of-the-art perceptual audio codecs using non waveform-preserving techniques such as bandwidth extension and joint stereo coding, known for presenting a challenge to existing quality predictors.

Results suggest that the derived distortion measure can be incorporated as an extension to existing automated perceptual quality assessment algorithms for improving prediction on spatially coded audio signals. 
\end{abstract}
\section{Introduction}
\label{sec:intro}

Since the advent of perceptual audio coders, a considerable interest arose in developing algorithms that can predict audio quality of the coded signals without relying on extensive subjective listening tests to save time and resources. Algorithms performing a so-called objective assessment of quality on monaurally coded signals such as PEAQ \cite{PEAQ} or POLQA~\cite{POLQAcite}~are widely spread. However, their performance for signals coded with spatial audio techniques is still considered unsatisfactory \cite{kmpf2010standardization}. In addition, non-waveform preserving techniques such as  bandwidth extension (BWE) are also known for causing these algorithms to overestimate the quality loss \cite{Ulovec} since many of the features extracted for analysis assume waveform preserving conditions. Spatial audio and BWE techniques are predominantly used at low-bitrate audio coding (around 32~kbps per channel).

It is assumed that spatial audio content of more than two channels can be rendered to a binaural representation of the signals entering the left and the right ear by using sets of Head Related Transfer Functions and/or Binaural Room Impulse Responses \cite{kmpf2010standardization, FallerBCC2}. Most of the proposed extensions for binaural objective assessment of quality are based on well-known binaural auditory cues related to the human perception of sound localization and perceived auditory source width such as Inter-aural Level Differences (ILD), Inter-aural Time Differences (ITD) and Inter- aural Cross-Correlation (IACC) between signals entering the left and the right ear \cite{seo2013perceptual, kmpf2010standardization, flener2017assessment, takanen2012a}. In the context of objective quality evaluation, features are extracted based on these spatial cues from reference and test signals and a distance measure between the two is used as a distortion index. The consideration of these spatial cues and their related perceived distortions allowed for considerable progress in the context of spatial audio coding algorithm design \cite{FallerBCC2}. However, in the use case of predicting the overall spatial audio coding quality, the interaction of these cue distortions with each other and with monaural/timbral distortions (especially in non waveform-preserving cases) renders a complex scenario \cite{conetta2015part2} with varying results when using the features to predict a single quality score given by subjective quality tests such as MUSHRA \cite{MUSHRA}. Other alternative models have also been proposed \citep{SchaeferICASSP} in which the output of a binaural model is further processed by a clustering algorithm to identify the number of participating sources in the instantaneous auditory image and therefore is also an abstraction of the classical auditory cue distortion models. However, the model in \citep{SchaeferICASSP} is mostly focused on moving sources in space and its performance is also limited by the accuracy and tracking ability of the associated clustering algorithm. The number of added features to make this model usable is also high in comparison to other models. 

Objective audio quality measurement systems should also employ the fewest, mutually independent and most relevant extracted signal features as possible to avoid the risk of over-fitting given the limited amount of ground-truth data for mapping feature distortions to quality scores provided by listening tests \cite{PEAQ}. One of the most salient distortion characteristics reported in listening tests for spatially coded audio signals at low bitrates is described as a collapse of the stereo image towards the center position and channel cross-talk \cite{baumgarte2002why}. 

We propose a simple feature aiming to describe the deterioration in the perceived auditory stereo image based on the change in loudness at regions that share a common \textit{panning index} \cite{AvendanoPan}. That is, regions in time and frequency of a binaural signal that share the same intensity level ratio between left and right channels, therefore corresponding to a given perceived direction in the horizontal plane of the auditory image. 

The use of directional loudness measurements in the context of auditory scene analysis for audio rendering of complex virtual environments is also proposed in \cite{Tsingos2004}, whereas the current work is focused on overall spatial audio coding quality objective assessment. 

The perceived stereo image distortion will be reflected as changes on a \textit{directional loudness map} of a given granularity corresponding to the amount of panning index values to be evaluated as a parameter. 

\section{Method}
\label{sec:Method}

\begin{figure*}[htb]
  \makebox[\textwidth][c]{\centerline{\includegraphics[width=14cm]{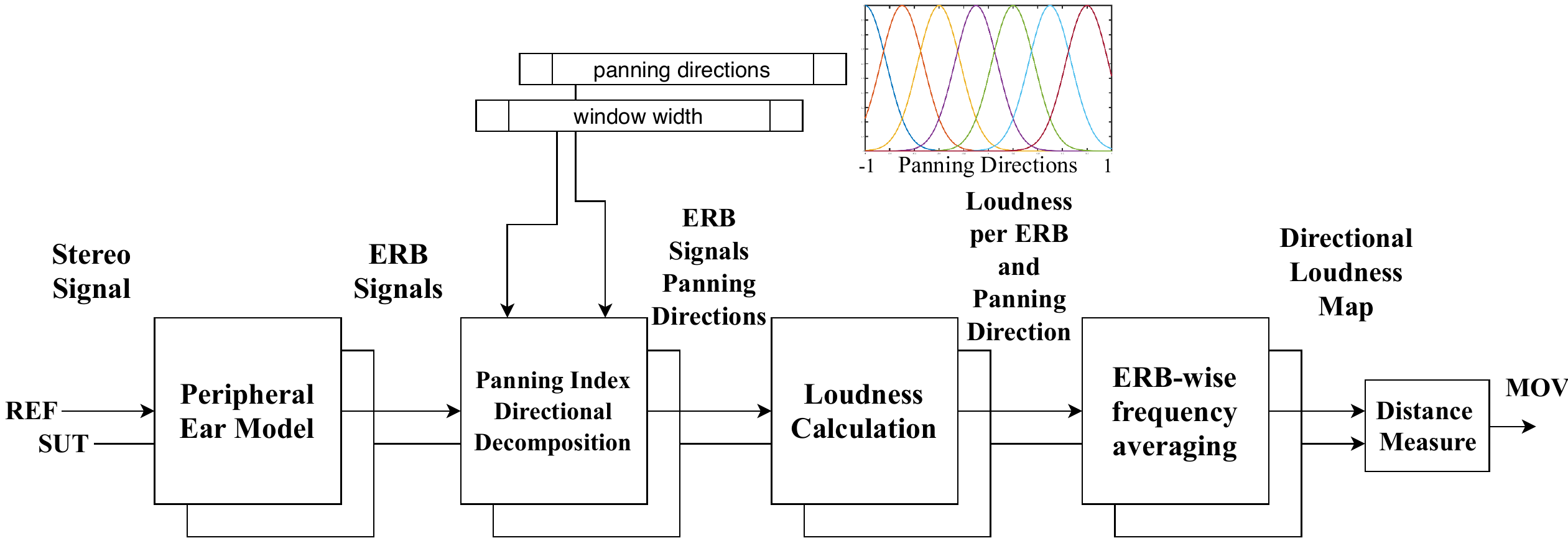}}}
\caption{Block Diagram for the proposed Model Output Value calculation. The panning directions and window width are represented as parameters given to the algorithm.}
\label{fig:block_diagram}
\end{figure*}

A two-channel reference signal (REF) and the two-channel signal under test (SUT) are processed in parallel in order to extract features that aim to describe -when compared- the perceived auditory quality degradation in the SUT. Multi-channel signals can be binaurally rendered to a two-channel signal as mentioned in the introduction. Figure \ref{fig:block_diagram} shows a block diagram for the proposed measure calculation. 

Both binaural signals are processed first by a peripheral ear model block. Each input signal is decomposed into the Short-Time Fourier Trasnform (STFT) domain (critically sampled) using a Hann window of block size $M=1024$ samples and overlap of $M/2$, giving a time resolution of 21 ms at a sampling rate of $F_s = 48$ kHz. The frequency bins of the transformed signal are then grouped to account for the frequency selectivity of the human cochlea in a total of $B=20$ bands with equal widths in the ERB scale \cite{MooreGlas}. Each band is then weighted by a value derived from the combined linear transfer function that models the outer and middle ear as explained in \cite{PEAQ}. 

The peripheral model outputs then signals $X_{i,b} (m,k)$ in each time frame $m$, and frequency bin $k$, and for each channel $i = \{L,R\}$ and each frequency group $b \in \{0,\ldots, B-1\}$, with different widths $K_b$ expressed in frequency bins.

\subsection{Directional Loudness Calculation}

The following concept is based on the method presented in \cite{AvendanoPan}, where a similarity measure between the left and right channels of a binaural signal in the STFT domain can be used to extract time and frequency regions occupied by each source in a stereophonic recording based on their designated panning coefficients during the mixing process.

Given the output of the peripheral model $X_{i,b} (m,k)$ a time-frequency (T/F) tile $Y_{i,b,\Psi_0}$  can be recovered from the input signal corresponding to a given panning direction $\Psi_0$ by multiplying the input by a window function $\Theta_{\Psi_0}$:
\label{ssec:DirLoud}

\begin{equation}
Y_{i,b,\Psi_0}(m,k) = X_{i,b} (m,k) \Theta_{\Psi_0} (m,k).
\end{equation}
The recovered signal will have the T/F components of the input that correspond to a panning direction $\Psi_0$ within a tolerance value. The windowing function can be defined as a Gaussian window centered at the desired panning direction:

\begin{equation}
\Theta_{\Psi_0} (m,k) = e^{-\frac{1}{2\xi} (\Psi(m,k) - \Psi_0)^2}
\end{equation}
where $\Psi(m,k)$ is the panning index as calculated in \cite{AvendanoPan} with a defined support of $[-1,1]$ corresponding to signals panned fully to the left or to the right, respectively. The signals $Y_{i,b,\Psi_0}$ will contain frequency bins whose values in the left and right channels will cause the function $\Psi$ to have a value of $\Psi_0$ or in its vicinity. All other components will be attenuated according to the Gaussian function. The value of $\xi$ represents the width of the window and therefore the mentioned vicinity for each panning direction. A value of $\xi = 0.006$ was chosen for a Signal to Interference Ratio (SIR) of $-60$~dB~\cite{AvendanoPan}. A set of $22$ equally spaced panning directions within $[-1,1]$ is chosen empirically for the values of $\Psi_0$. For each recovered signal, a loudness calculation \cite{ZwickerLoud} at each ERB band and dependent on the panning direction is expressed as:

\begin{equation}
L_{b, \Psi_0}(m) = \left(\frac{1}{K_b} \sum_{k \in b} Y_{DM_{b,\Psi_0}}(m,k)^2\right)^{0.25}
\end{equation}
where $Y_{DM}$ is the sum signal of channels $i = \{L,R\}$. The loudness is then averaged over all ERB bands to provide a directional loudness map  defined over the panning domain $\Psi_0 \in [-1,1]$ over time frame $m$:

\begin{equation}
\label{eq:loudness_map}
L(m, \Psi_0) = \frac{1}{B}\sum_{\forall b} L_{b,\Psi_0}(m).
\end{equation}

For further refinement Equation \ref{eq:loudness_map} can be calculated only considering a subset of the ERB bands corresponding to frequency regions of $1.5$ kHz and above to accommodate to the sensitivity of the human auditory system to level differences in this region, according to the \textit{duplex theory} \cite{duplextheo}. In this work, bands $b \in \{7,\ldots, 19\}$ were used corresponding to frequencies from $1.34$ kHz to $F_s /2$.

As a final step, directional loudness maps for the duration of the reference signal and SUT are subtracted and the absolute value of the residual is then averaged over all panning directions and time producing a single number termed Model Output Variable (MOV), following the terminology in \cite{PEAQ}. This number effectively expressing the distortion between directional loudness maps of reference and SUT, is expected to be a predictor of the associated subjective quality degradation reported in listening tests. 

\begin{figure*}[htb]
\begin{center}

\begin{minipage}[b]{\imgsize \linewidth}
  \makebox[\textwidth][c]{\centering{\includegraphics[width=\imgs2]{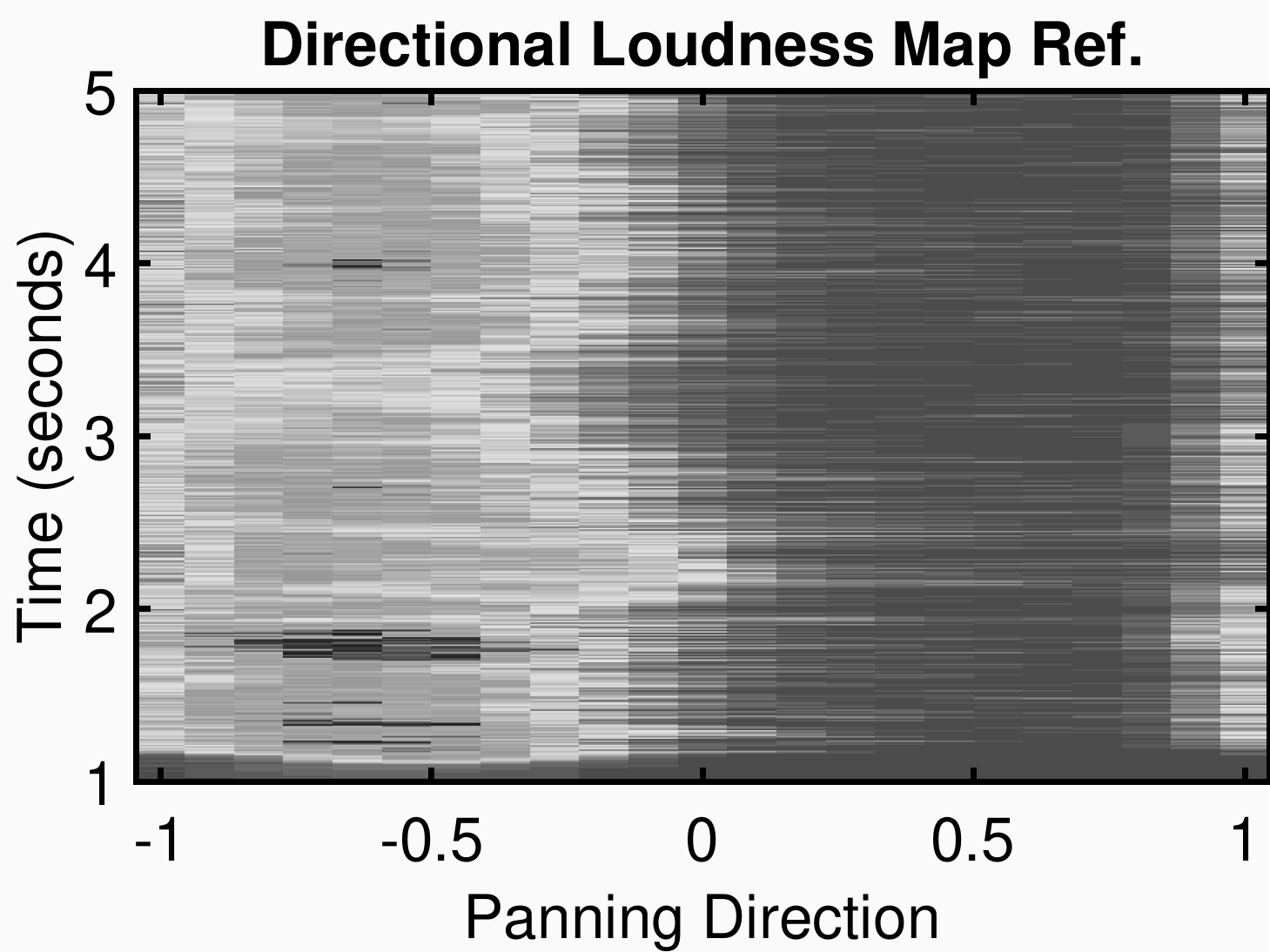}}}
  \vspace{0.05cm}
  \centering{(a) Directional Loudness Map REF  }\medskip
\end{minipage}
\hfill
\begin{minipage}[b]{\imgsize \linewidth}
  \makebox[\textwidth][c]{\centering{\includegraphics[width=\imgs2]{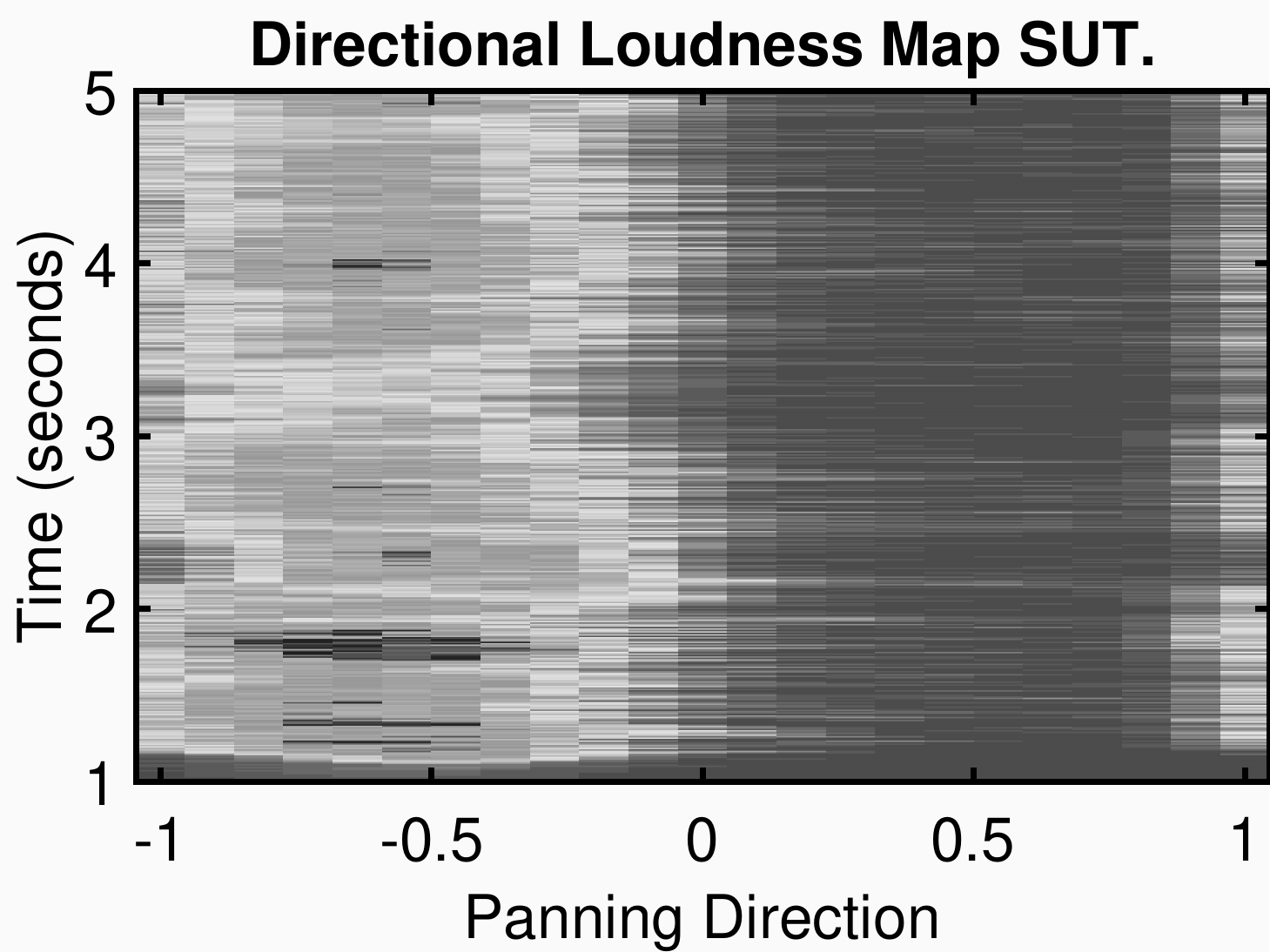}}}
  \vspace{0.05cm}
  \centering{(b) Directional Loudness Map SUT}\medskip
\end{minipage}
\hfill
\begin{minipage}[b]{\imgsize \linewidth}
  \makebox[\textwidth][c]{\centering{\includegraphics[width=\imgs2]{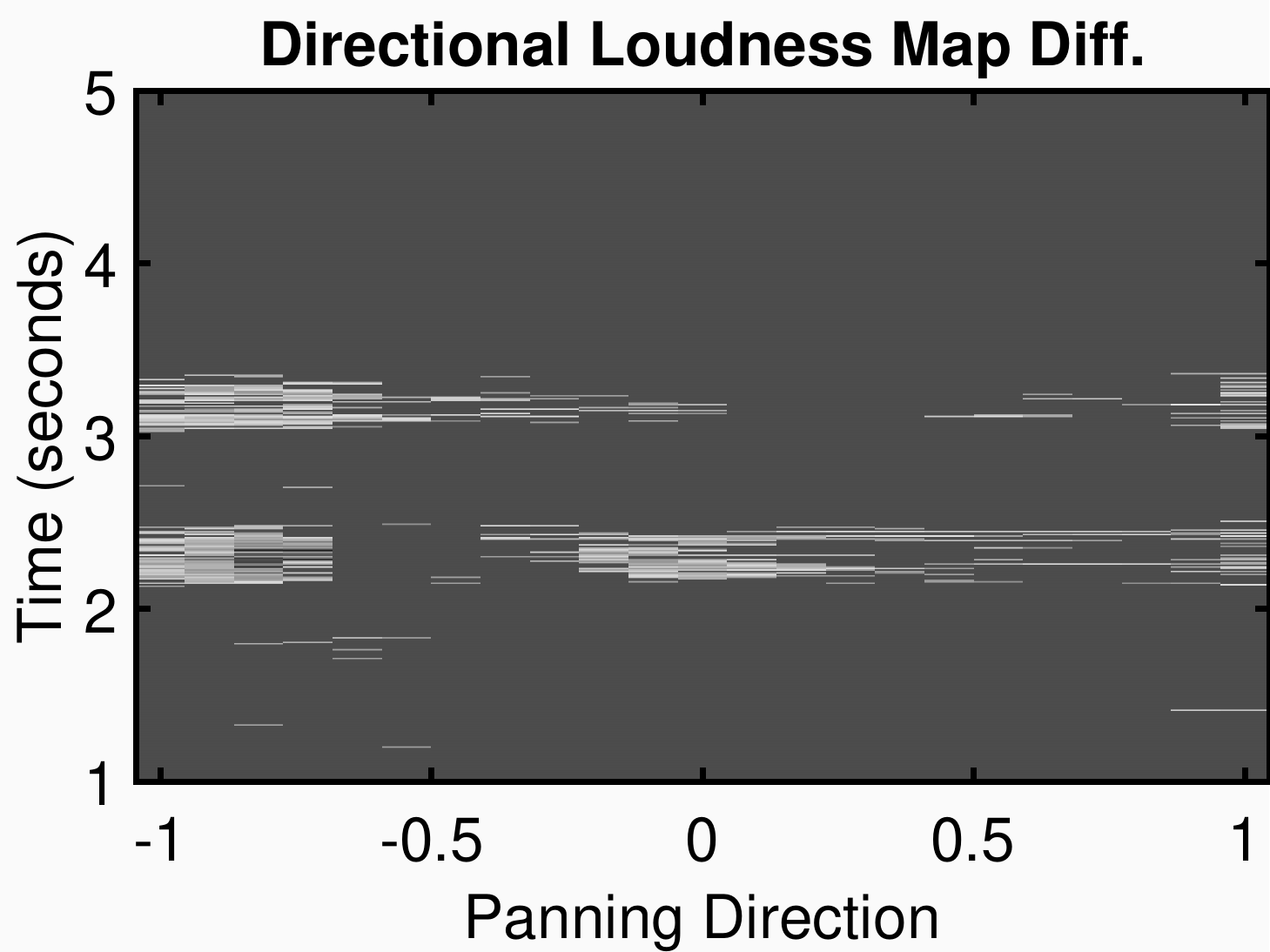}}}
  \vspace{0.05cm}
  \centering{(c) Directional Loudness Map DIFF}\medskip
\end{minipage}

\caption{Example of a solo violin recording of 5 seconds of duration panned to the left. Clearer regions on the maps represent louder content.  The degraded signal (SUT) presents a temporal collapse of the panning direction of the auditory event from left to center between times 2-2.5 sec and again at 3-3.5 sec.}

\end{center}

\label{fig:dir_loudness}

\end{figure*}

Figure \ref{fig:dir_loudness} shows an example of application of the concept of a directional loudness map to a pair of reference (REF) and degraded (SUT) signals, and the absolute value of their difference (DIFF).

\section{Experiment description}
\label{sec:Experiment}

In order to test and validate the usefulness of the proposed MOV, a regression experiment similar to the one in \cite{delgado2018energy} was carried out in which MOVs were calculated for reference and SUT pairs in a database and compared to their respective subjective quality scores from a listening test. The prediction performance of the system making use of this MOV is evaluated in terms of correlation against subjective data ($R$), absolute error score ($AES$), and number of outliers ($\nu$), as described in \cite{PEAQ}.

The database used for the experiment corresponds to a part of the Unified Speech and Audio Coding (USAC) Verification Test \cite{USACdatabase} Set 2, which contains stereo signals coded at bitrates ranging from 16 to 24 kbps using joint stereo \cite{baumgarte2002why} and bandwidth extension tools along with their quality score on the MUSHRA scale. Speech items were excluded since the proposed MOV is not expected to describe the main cause of distortion on speech signals. A total of 88 items (average length 8 seconds) remained in the database for the experiment.

To account for possible monaural/timbral distortions in the database, the outputs of an implementation of the standard PEAQ (Advanced Version) termed Objective Difference Grade (ODG) and POLQA, named Mean Opinion Score (MOS) were taken as additional MOVs complementing the directional loudness distortion (DLD) described in the previous section. All MOVs were normalized and adapted to give a score of 0 for indicating best quality and 1 for worst possible quality. Listening test scores were scaled accordingly. 

A a fraction of the database (60\%, 53 items) was reserved for training a regression model using Multivariate Adaptive Regression Splines (MARS) \cite{flener2017assessment} mapping the MOVs to the items subjective scores. The remainder (35 items) were used for testing the performance of the trained regression model. In order to remove the influence of the training procedure from the overall MOV performance analysis, the training/testing cycle was carried out 500 times with randomized training/test items and mean values for $R$, $AES$, and $\nu$ were considered as performance measures. 
 
\section{Results and discussion}
\label{sec:Results}

\begin{table}[t]  
  \centering
    \resizebox{\columnwidth}{!}{%
      \begin{tabular}{l|c|c|c|c}
        \textbf{MOV Set (N)} & \textbf{R} &\textbf{AES} & $\mathbf{\nu}$ & $\mathbf{R^2_{a}} $ \\ \hline
        MOS + ODG (2) & 0.77 & 2.63 & 12 & 0.05 \\ 
        MOS + ODG + CHOI (5) & 0.77 & 2.39 & 11 & 0.50 \\ 
        MOS + ODG + EITDD (3) & 0.82 & 2.0 & 11 & 0.02 \\
        MOS + ODG + SEO (6) & 0.88 & 1.65 & 7 & 0.52 \\ 
        MOS + ODG + DLD (3) & 0.88 & 1.69 & 8 & 0.70 \\ 
      \end{tabular}
    }
				\caption{Mean performance values for 500 training/testing cycles of the regression model with different sets of MOVs. CHOI represents the 3 binaural MOVs as calculated in \cite{choi2008objective}, EITDD corresponds to the high frequency envelope ITD distortion MOV as calculated in \cite{seo2013perceptual}. SEO corresponds to the 4 binaural MOVs from\cite{seo2013perceptual}, including EITDD. DLD is the proposed MOV. The number in parenthesis represents the total number of MOVs used.}
	\label{tab:results}
\end{table}

Table \ref{tab:results} shows the mean performance values (correlation, absolute error score, number of outliers) for the experiment described in Section \ref{sec:Experiment}. In addition to the proposed MOV, the methods for objective evaluation of spatially coded audio signals proposed in \cite{choi2008objective} and \cite{seo2013perceptual} were also tested for comparison. Both compared implementations make use of the classical inter-aural cue \textit{distortions} mentioned in the introduction: IACC distortion (IACCD), ILD distortion (ILDD), and ITDD. 

As mentioned, the baseline performance is given by ODG and MOS, both achieve $R=0.66$ independently but present a combined performance of $R=0.77$. This confirms that the features are complimentary in the evaluation of monaural distortions. 

Considering the work of Choi et. al. \cite{choi2008objective}, the addition of the three binaural distortions (CHOI in Table \ref{tab:results}) to the two monaural quality indicators (making up to five joint MOVs) does not provide any further gain to the system in terms of prediction performance for the used dataset. 

In \cite{seo2013perceptual}, some further model refinements were made to the mentioned features in terms of lateral plane localization and cue distortion detectability. In addition, a novel MOV that considers high frequency envelope inter-aural time difference distortions (EITDD) \citep{EITDD_theory} was incorporated. The set of these four binaural MOVs (marked as SEO in Table \ref{tab:results}) plus the two monaural descriptors (6 MOVs in total) significantly improves the system performance for the current data set. 

Looking at the contribution in improvement from EITDD suggests that frequency time-energy envelopes as used in joint stereo techniques \cite{baumgarte2002why} represent a salient aspect of the overall quality perception.

The presented MOV based on directional loudness map distortions (DLD) correlates even better with the perceived quality degradation than EITDD, reaching similar performance as the combination of all the binaural MOVs of \cite{seo2013perceptual}, while using one additional MOV to the two monaural quality descriptors, instead of four. Using fewer features for the same performance will reduce the risk of over-fitting and indicates their higher perceptual relevance. This can be confirmed by observing the adjusted R squared values $\mathbf{R^2_{a}}$ for each model, reflects prediction power of the model while penalizing model complexity \cite{Efron_R_adj}.

A maximum mean correlation $R=0.88$ shows that there is still room for improvement. One important limitation of the current model on which the proposed feature is based is that it assumes a simplified description of stereo signals in which auditory objects are only localized in the lateral plane by means of ILDs, which is usually the case in studio-produced audio content \cite{AvendanoPan}. For ITD distortions usually present when coding multi-microphone recordings or more natural sounds, the model needs to be either extended or complemented by a suitable ITD distortion measure.

\section{Conclusions and future work}

A distortion metric was introduced describing changes in a representation of the auditory scene based on loudness of events corresponding to a given panning direction. The important increase in performance compared to the monaural-only quality prediction shows the effectiveness of the proposed method. The approach also suggests a possible alternative or complement in quality measurement for low bitrate spatial audio coding where established distortion measurements based on classical binaural cues do not perform satisfactorily, possibly due to the non-waveform preserving nature of the audio processing involved. 

The performance measurements show that there are still areas for improvement towards a more complete model that also includes auditory distortions based on effects other than channel level differences. Future work also includes studying how the model can describe temporal instabilities/modulations in the stereo image as reported in \cite{baumgarte2002why} in contrast to static distortions. 

\clearpage
\vfill\pagebreak

\bibliographystyle{IEEEbib}
\input{where_is_the_bibfile.tex}
\end{document}

%% file: where_is_the_bibfile.tex
\bibliography{./papers_st_polqa}